\documentclass[12pt,preprint]{aastex}

\newcommand{\Msolar}{\mbox{\,M$_\odot$}}        
\newcommand{\kms}{\mbox{\,km\,s$^{-1}$}}                           
\newcommand{\degrees}{$^{\circ}$}

\newcommand{\um}{$\mu$m}
\newcommand{\xco}{$^{13}$CO}
\newcommand{\co}{$^{12}$CO}

\slugcomment{This draft \today}
\shorttitle{The Warm Shell in Perseus}
\shortauthors{Ridge et al.}

\defcitealias{cernis03}{CS}
\defcitealias{gbmm90}{GBMM}

\begin{document}
\title{The COMPLETE Nature of the Warm Dust Shell in Perseus}


\author{Naomi A. Ridge, Scott L. Schnee, Alyssa A. Goodman, Jonathan B. Foster}
\affil{Harvard-Smithsonian Center for Astrophysics, 
60 Garden St., Cambridge, MA, 02138, USA}
\email{NAR: nridge@cfa.harvard.edu, SLS: sschnee@cfa.harvard.edu, AAG: agoodman@cfa.harvard.edu}

\begin{abstract}
The Perseus molecular cloud complex is a $\ga$30\,pc long chain of
molecular clouds most well-known for the two star-forming clusters
NGC\,1333 and IC\,348 and the well-studied outflow source in
B5. However, when studied at mid- to far-infrared wavelengths the
region is dominated by a $\sim$10\,pc diameter shell of warm dust,
likely generated by an H{\sc ii} region caused by the early B-star
HD\,278942. Using a revised calibration technique the COMPLETE team
has produced high-sensitivity temperature and column-density maps of
the Perseus region from IRAS Sky Survey Atlas (ISSA) 60 and 100\,\um\
data.  In this paper, we combine the ISSA based dust-emission maps
with other observations collected as part of the
COMPLETE\footnote{{\bf C}o{\bf O}rdinated {\bf M}olecular {\bf P}robe
{\bf L}ine, {\bf E}xtinction and {\bf T}hermal {\bf E}mission;
http://cfa-www.harvard.edu/COMPLETE} Survey, along with archival
H$\alpha$ and MSX observations.  Molecular line observations from
FCRAO and extinction maps constructed by applying the NICER method to
the 2MASS catalog provide independent estimates of the ``true''
column-density of the shell. H$\alpha$ emission in the region of the
shell confirms that it is most likely an H{\sc ii} region located
behind the cloud complex, and 8\,\um\ data from MSX indicates that the
shell may be interacting with the cloud.  Finally, the two
polarisation components seen towards background stars in the region by
\citet{gbmm90} can be explained by the association of the stronger
component with the shell. If confirmed, this would be the first
observation of a parsec-scale swept-up magnetic field.

\end{abstract}


\section{Introduction}
\label{intro}
The interaction of newly formed stars with their parent cloud,
particularly in the case of massive stars, can have significant
consequences for the subsequent development of the cloud. Hot, massive
stars will cause cloud disruption through their ionizing flux as well
as through the momentum in their stellar winds. Alternatively, given
the right conditions in the cloud, stellar winds may also trigger the
collapse of cloud cores and induce further star formation.

One region where a stellar wind may be triggering star formation is
the Perseus Molecular Cloud complex, a well-studied chain of molecular
clouds, extending approximately 30\,pc in length.  The complex has a
total mass of $\sim$1.3$\times$10$^4$\Msolar, assuming it is at a
distance of 260\,pc (although actual distance estimates range from
230\,pc
\citep{cernis90} to 350\,pc
\citep{hj83}, and it is clear from the large ($\ga$ 10\,\kms) velocity
range of molecular gas that a single distance to the entire complex is
unlikely). Star formation is ongoing in several parts of the complex,
most obviously around the two reflection nebulae
\objectname{IC\,348} and \objectname{NGC\,1333}.  Several surveys
\citep[e.g.][]{llm93,asr94,ll95,luhman} have established that there is a
population of pre-main sequence stars located both within the clusters
and throughout the complex, but relatively few high-mass stars have
been found.  

The Perseus complex is not just interesting for its star-forming
properties. An almost complete shell of enhanced emission can be seen
in IRAS data toward the center of the molecular cloud complex (figure
\ref{ext_temp}).  The shell, referred to in previous papers as
\objectname{\hbox{G\,159.6$-$18.5},} has a diameter of 0.75\degrees, or
10\,pc at the distance of the molecular cloud. Although its existence
has been known for about 15 years (it was first described by
\citet{ps89} in a conference proceedings and further discussed by
\citet{fpjd94}) there has been little investigation into the nature
or source of the shell, and it has been mostly ignored by studies of
star-formation in the region. Based on radio data, \citeauthor{fpjd94}
argued that the feature is caused by a supernova remnant, which if
true, would be one of the closest to the Sun, and have the highest
known Galactic latitude for such an object. More recently \citet{dZ99}
associated the shell with the B star HD\,278942, located at its
geometric center. \citet{ander2000} performed a multi-wavelength study
of the star and its surroundings. They reclassified HD\,278942 as an
O9.5-B0 V star with an age of 8\,Myr, and found weak radio continuum
emission with a flat spectral index, consistent with an optically thin
H{\sc ii} region filling the shell. However, partly due to the lack of
short-spacing information in their interferometer maps leading to
large uncertainties on the derived radio-continuum fluxes, they could
not rule out a negative spectral index, leaving the possibility that
the radio continuum is due to synchrotron emission from the
interaction between a stellar wind or supernova remnant and the
molecular cloud.
\citeauthor{ander2000}'s other observations concentrated on the
central star HD\,278942, and hence provide little insight into the
nature of shell itself.

The Perseus molecular cloud complex is a target of the Spitzer Space
Telescope (SST) Legacy program ``From Molecular Cores to Planet
Forming Disks''
\citep[hereafter c2d;][]{evans03}, while the two clusters NGC\,1333
and IC\,348 are included in an SST Guaranteed Time Observer (GTO)
program. These programs aim to determine the distribution of young
stars and clusters, and investigate their association with known dense
cores. The ongoing COMPLETE Survey \citep{data}, is a large
international effort which has coordinated its observations with
c2d.  COMPLETE aims to obtain and compare high quality molecular line,
submillimeter continuum, far-infrared column-density and near-infrared
extinction data over the extents of three of the star-forming
molecular clouds targeted by c2d, including the whole of Perseus. 

As part of COMPLETE, we have recently used 60 and 100\,\um\ IRAS maps
of the Perseus region to create new high-sensitivity temperature and
column-density maps
\citep{schnee_iras}. Here we intercompare
the recalibrated far-IR results with COMPLETE near-infrared extinction
and \xco\ maps of the same region
\citep{data}, an H$\alpha$ image from the VTSS\footnote{Virginia Tech
Spectral Line Survey; http://www.phys.vt.edu/~halpha/} Survey, and the
MSX 8\,\um\ image. A detailed picture of the shell which, would not be
possible by looking at each data set alone emerges and suggests a
complex picture of Perseus as a group of cold molecular clouds being
impacted gently from behind by a stellar wind bubble.

\section{Data and Analysis}
\subsection{IRAS Temperature and Column-Density Maps}
\label{temp}
Figure \ref{ext_temp} shows maps of colour temperature and
column-density in Perseus from \citet{schnee_iras}. These were
constructed from recalibrated 60 and 100\,\um\ IRIS\footnote{Improved
Recalibration of the {\it IRAS} Survey} images \citep{iris}. These
data provide excellent correction for the effects of zodiacal dust and
striping in the IRAS images and improved gain, offset and zero-point
calibration over earlier releases of the IRAS data
\citep{schnee_iras}. Near-infrared 
extinction maps (see section \ref{extinction}) were used to constrain
the conversion factor between 100\um\ optical depth and visual
extinction (column density), as using the ``standard'' values for this
conversion \citep[e.g.\ ][]{wmd94} results in a significant
miscalculation of the extinction \citep{schnee_iras}. Hence these maps
provide the best dust-temperature and extinction maps constructed from
IRAS data to date.

The positions of the well-known star-forming clusters NGC\,1333 and
IC\,348 are indicated in figure \ref{ext_temp}.  The shell is clearly
seen as an enhancement in column density, and is filled with warm
material, evident in the temperature map.  Although on careful
inspection, the known dark cores in Perseus are visible in the map,
the general morphology (dominated by the shell) is very different from
extinction or molecular line maps of the region which are dominated by
the chain of clouds (see sections
\ref{extinction} and \ref{co_sect}).
%

\begin{figure}
\rotate{
\caption{Column-density 
in units of A$_V$ derived from IRAS 60 and 100\um emission, (left) and
temperature overlaid with contours of column-density (right), in the
direction of the Perseus molecular cloud complex.  The position of the
well-known star-forming clusters, NGC\,1333 and IC\,348 are shown for
orientation.  A bright shell of emission, filled with warm material
dominates the region at these wavelengths. The cross indicates the
position of the B-star HD\,278942, which has been proposed as the
progenitor of the shell.  THIS FIGURE SHOULD BE FULL-PAGE/LANDSCAPE
ORIENTATION.
\notetoeditor{This figure should be typeset as a full page in 
landscape orientation}
\label{ext_temp}}}
\end{figure}

%
%

 
\subsection{Extinction from Near-Infrared Colour Excess}
\label{extinction}
As part of COMPLETE, we have constructed a map of the extinction
towards Perseus with comparable resolution (5$'$) to the 4.7$'$
IRAS-based maps, by applying the
NICER\footnote{Near-Infrared Colour Excess Revisited.} algorithm
\citep{la01} to data from the Two Micron All-Sky Survey \citep[2MASS; ][]
{data}.  This
method uses the near-infrared colour excess of background stars to
determine reddening along a line of sight, and hence does not rely on
assumptions about grain size distribution or emissivity. The
extinction determined in this way should therefore provide the best
measure of the true column density distribution of material
\citep*{grs}.The 2MASS/NICER extinction map is shown in figure \ref{2mass}.
Here the more familiar view of Perseus, as a chain of molecular
clouds, from \objectname{B5} in the northeast, through IC348, B1 and
NGC1333 in the southwest are clearly visible. Again the position of
HD278942 is indicated by a cross, and although faint, the northern
half of the shell, centered on the star, can be made out in the image.
\begin{figure}
\caption{Color: Near-IR based extinction map of Perseus, covering the same area shown in figure \ref{ext_temp}. This map was constructed by applying the NICER algorithm to data from the 2MASS catalog. 
Again the cross indicates the position of HD\,278942. Contours: IRAS
column-density (levels are A$_V$=2.5,3.5).  From
\citet{data}.
\notetoeditor{Should be typeset to appear the same size as each panel in 
figure 1.}
\label{2mass}}
\end{figure}

\subsection{Molecular line Emission from \xco}
\label{co_sect}
A map of \xco\ integrated intensity (representing gas column density)
with 44$''$ resolution, obtained for COMPLETE at the Five College
Radio Astronomy Observatory (FCRAO) 14m Telescope, is shown in figure
\ref{co}. A full description of the data acquisition and reduction
methods is given in \citet{data}.

Although not as extensive as the IRAS and 2MASS/NICER images, the
higer resolution\footnote{Figure reproductions presented here do not
aloow for the reader to see the full detail of the map -- higher
resolution versions and the raw data are available at
http://cfa-www.harvard.edu/COMPLETE} of this map reveals significant
substructure when compared to those images. Like the 2MASS/NICER
extinction map, the
\xco\ integrated intensity shows the familiar chain of molecular
clouds, with no hint of the existence of the warm dust shell so clear
in the IRAS-based image.

\begin{figure}
\caption{
The ``traditional'' picture of Perseus -- a chain of molecular clouds
shown here in \xco\ integrated intensity. The solid black line shows
the border of the observed region. The cross indicates the position of
HD\,278942, and the locations of well-known star-forming sites are
indicated.  This map does not extend significantly into the area of
the shell (shown by the dashed circle), but there is no hint of its
existence from the \xco\ integrated emission.  From
\citet{data}.
\notetoeditor{Should be typeset across two columns}
\label{co}}
\end{figure}

\subsection{H$\alpha$ image from VTSS}

Figure \ref{halpha_iras} shows a 3\degrees$\times$3\degrees\ H$\alpha$
image centered on $\alpha$=03$^{\rm h}$39$^{\rm m}$57\fs1,
$\delta$=+31\degrees55$^{\rm m}$15$^{\rm s}$ obtained via {\it
Skyview} from the VTSS, overlaid with contours of IRAS column-density.
The image has a resolution of 6$'$.

There is clearly ionised gas filling the shell, indicating that it is
likely to be an H$ii$ region or stellar wind. The H$\alpha$ emission
displays three distinct peaks, the brightest to the northeast. The
substructure in the H$\alpha$ emission, in particular tha apparent
lack of emission to the southeast, will be discussed in section
\ref{twodust}.
\begin{figure}
\caption{
H$\alpha$ emission \citep{fink04,vtss}, overlaid with contours of
N$_H$ (in units of A$_V$) from IRAS. H$\alpha$ emission fills the
shell, indicating that it is likely to be an H{\sc ii} region or
stellar wind.
\label{halpha_iras}}
\end{figure}

\subsection{Mid-infrared observations from MSX}

The shell was the subject of a pointed observation by the Midcourse
Space Experiment (MSX) satellite, which provided images at four
mid-infrared wavebands. The shell was clearly detected at 8.3\um (as
shown in figure \ref{msx}) and 12.1\,\um\
\citep{kraemer}. It was not detected in the 14.7 and 21\,\um\ bands,
but this is most likely due to the lower sensitivity of those bands,
as it is clearly present in the 12 and 25\,\um\ bands of IRAS and has
been seen by both the IRAC and MIPs instruments on the Spitzer Space
telescope (K. Stapelfeldt, J. Jorgensen, personal communication). The
higher resolution of the 8\um\ MSX observations compared to IRAS
reveals complex filamentary structures within the shell
\citep{kraemer}. In particular, there are 3 bright knots of 8\um\
emission, labelled A, B and C in figure \ref{msx}, which also
correspond to enhancements in the extinction at those positions.

\begin{figure}
\caption{MSX 8\,\um\ image of the shell 
overlaid with contours of extinction (from 2MASS/NICER). Contour
levels start at an A$_V$ of 3 and increase in increments of 2. The
bright 8\,\um\ ``knots'', indicated by red triangles and labelled A,
B and C are spatially coincident with enhancements in the extinction.
\xco\ spectra at these positions are shown in figure \ref{co_spec} and discussed in section \ref{impact}.
\label{msx}}
\end{figure}

\section{Discussion}
\label{discuss}

\subsection{Could HD\,278942 be the driving source of the shell?}
The fact that the shell is visible in the 2MASS extinction map and is
filled with preferentially heated dust emission tells us that it is a
bubble of heated material, enclosed in a colder shell of increased
density (visible only on the edges). Based on multiwavelength data,
\citet{ander2000} suggested that the shell was the result of an
expanding stellar wind from HD\,278942. This is supported by the fact
that the shell is filled with H$\alpha$ emission, as is shown in
figure
\ref{halpha_iras}.

Historically, the combined effects of the shell and the high reddening
towards HD\,278942 (A$_V$ = 7.4\,mag; \citealt{ander2000}) has made
spectral classification of the star problematic, with values in the
literature ranging from F2 (Hipparcos) through B3\,III
\citep{cernis93} to O9.5$\rightarrow$B0\,V
\citep{ander2000}. The situation is further complicated by the
possibility that HD\,278942 may be a photometric B3\,III + F5\,I
binary \citep{cernis93}. Recent spectra obtained by \citet{steen}
confirm that HD\,278942 is most likely an early--mid B-star. Assuming
a local density, n$_H$, of 1\,cm$^{-3}$, such a star could
produce an H{\sc ii} region of several parsecs in size
\citep{osterbrock}. Given the uncertainty in the star's spectral
classification and the local density, it is therefore plausible that
HD\,278942 created the ~10\,pc diameter shell.

\subsection{Evidence for Two Dust Populations in Perseus}
\label{twodust}

In order to compare the properties of the shell with other regions of
the cloud, we selected by eye an annular region of the IRAS
column-density image which contains all of the shell emission.  Figure
\ref{annulus} shows this region, as a hatched annulus overlaid on the
IRAS column density image. In the remainder of the text we will refer
to the hatched region of the annulus as the ``shell component'', the
circular region contained within the annulus as ``inside shell'' and
all points in the image that are outside of the annulus as the ``cloud
component''. Also shown are two small grey boxes, which indicate the
positions of the two young clusters NGC\,1333 and IC\,348 which we
excluded from our analysis.

\begin{figure}
\caption{IRAS extinction map of Perseus. The hatched area
indicates the region we define as the ``shell component'', points
within the circular region contained within the annulus as ``inside
shell'' and points outside the outer circle are defined as the ``cloud
component.''  The two grey boxes show the small regions around IC\,348
(left) and NGC\,1333 (right) which were excluded from the analysis
in section \ref{twodust}.
\label{annulus}}
\end{figure}

Our new IRAS calibration technique \citep{schnee_iras} leads to
temperature and column-density maps which look somewhat different to
those presented in \citet{ander2000}. They interpreted their
temperature map as showing that the ring is at a temperature minimum
of 24\,K with the dust interior to the shell at a higher
temperature. The fact that they also found the cloud material outside
the shell marginally warmer than the shell itself (making the shell a
temperature minimum) is probably an artifact of the way they created
their temperature map.

\begin{figure}
\plotone{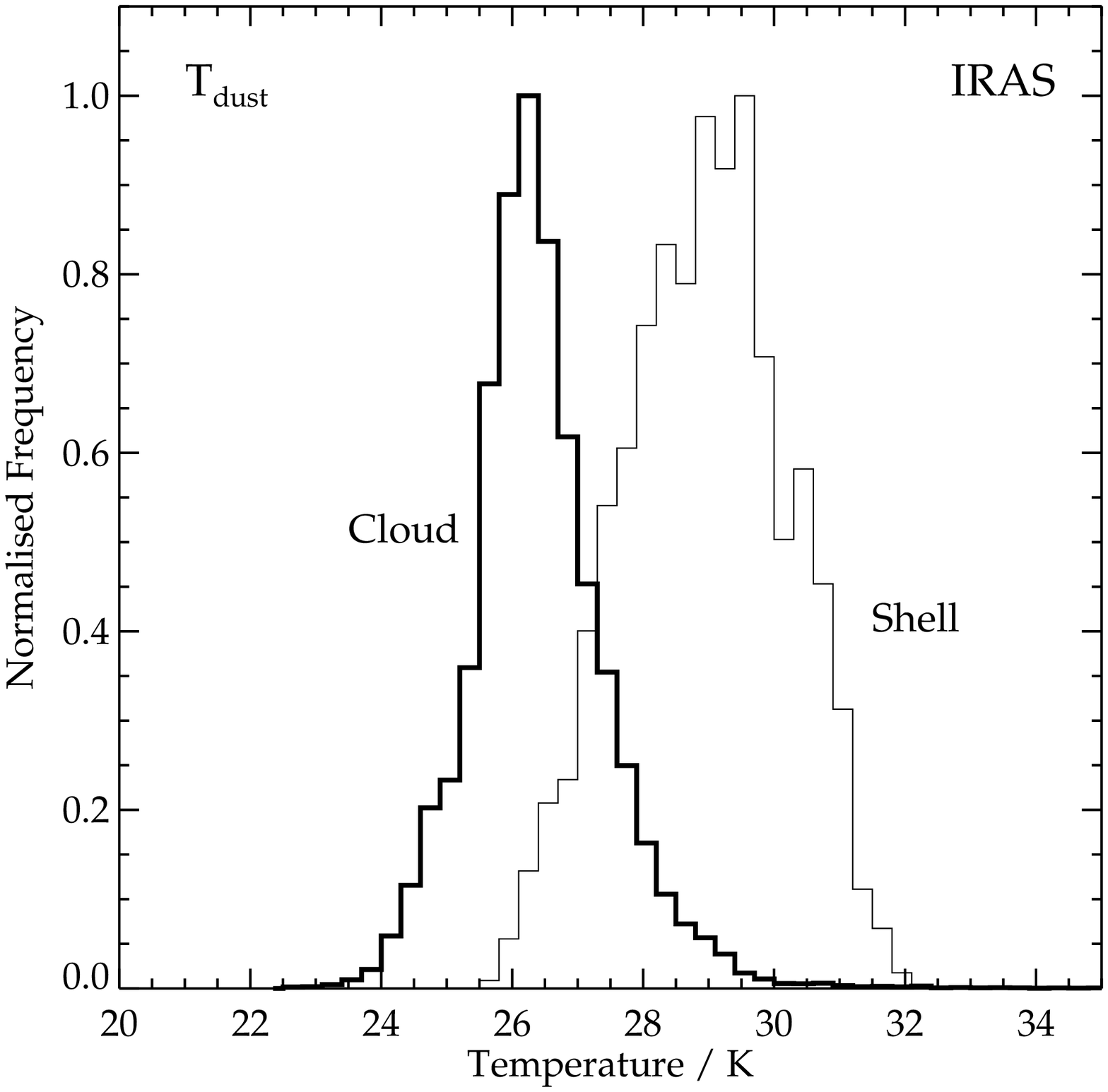}
\caption{
Histogram of dust temperature determined from IRAS 60 and 100\um\
emission. The thick line represents the cloud component, while the
thin line shows the shell component. Note that small areas containing
the young clusters IC\,348 and NGC\,1333 were excluded
(see fig. \ref{annulus}). The histograms are normalised due to the
large difference in the number of points in the two components (65000
points for the cloud versus 6800 for the shell component).
\label{temp_hist}}
\end{figure}

We find an average temperature in the shell component of 28.9\,K, with
a maximum of 32\,K.  The distribution of temperatures within the shell
component and cloud component are shown in figure
\ref{temp_hist}. It is clear that the shell component is significantly
warmer than the molecular cloud material which surrounds it. This is
supported by a two-sided Kolmogorov-Smirnov (KS) test, which gives a
probability of $<10^{-5}$ that the two samples are drawn from the same
population.

In figure \ref{ext_temp_scatter} we show a scatter plot of the colour
temperature versus column-density derived from recalibrated IRAS 60
and 100\,\um\ emission \citep{schnee_iras}. Three distinct populations
are seen in the emission, indicated by the green points (cloud
component), red points (shell component) and blue points (inside
shell). Clearly the shell and its interior are warmer and denser on
average than the surrounding cloud.

The difference in density can be seen more clearly in the histograms
shown in the left panel of figure \ref{ext_hist}. The cloud component
has a mean extinction of 1.6\.mag., while the mean of extinction in
the shell component is almost double that (3.1\,mag.)  Again, a
two-sided KS test gives a probability of $\ll 10^{-5}$ that the two
populations (cloud and shell) are drawn from the same parent
population.

\begin{figure}
\caption{Column density vs. temperature derived
from IRAS 60 and 100\,\um\ emission. Green points indicate cloud
component points (i.e. outside the outer circle of the annulus shown
in figure \ref{annulus}).  Red dots indicate points within the hatched
annulus (shell component) and blue dots
indicate points from the interior of the shell (i.e. inside the inner circle of
the annulus).
\label{ext_temp_scatter}}
\end{figure}
\begin{figure}
\plottwo{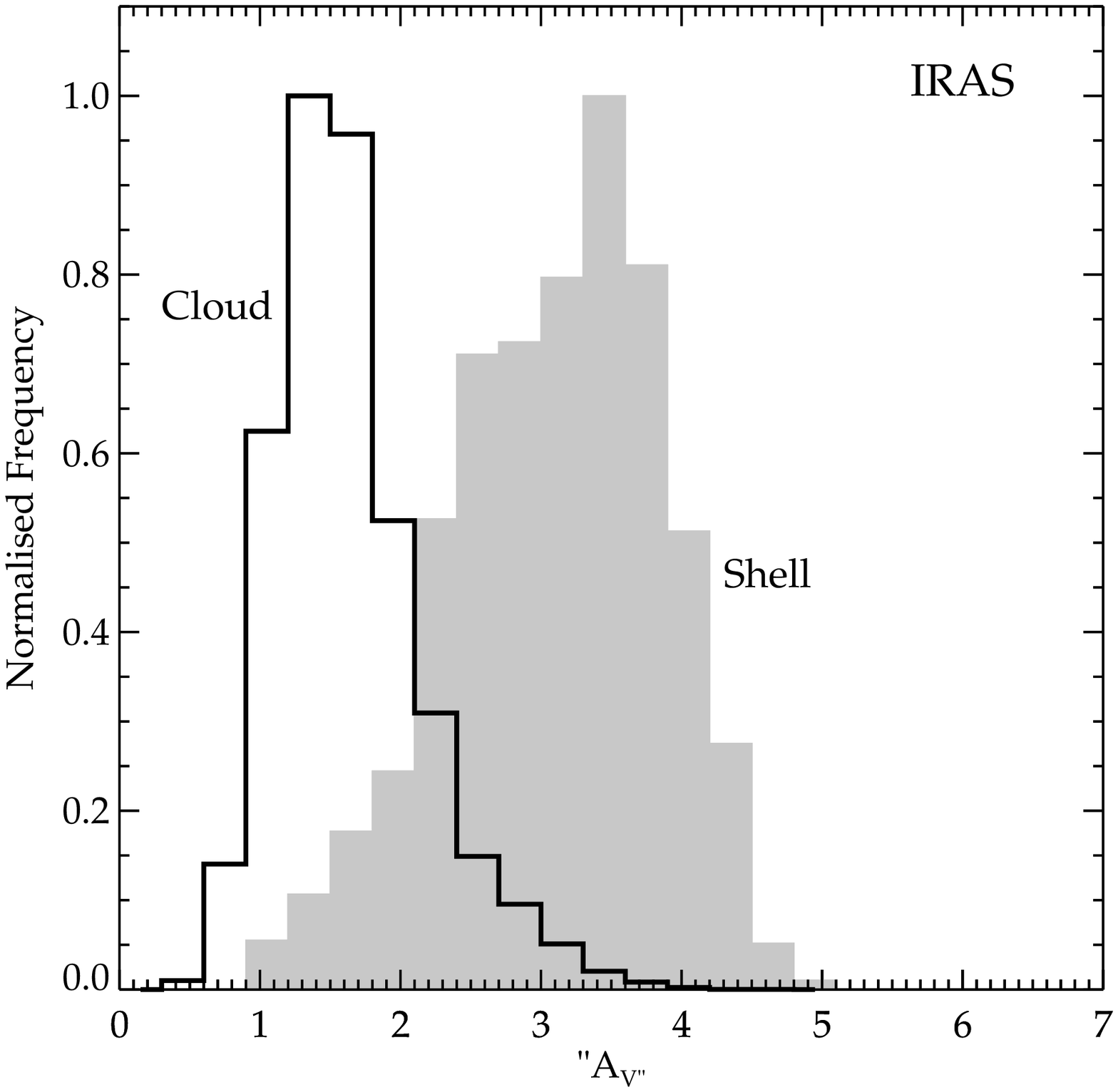}{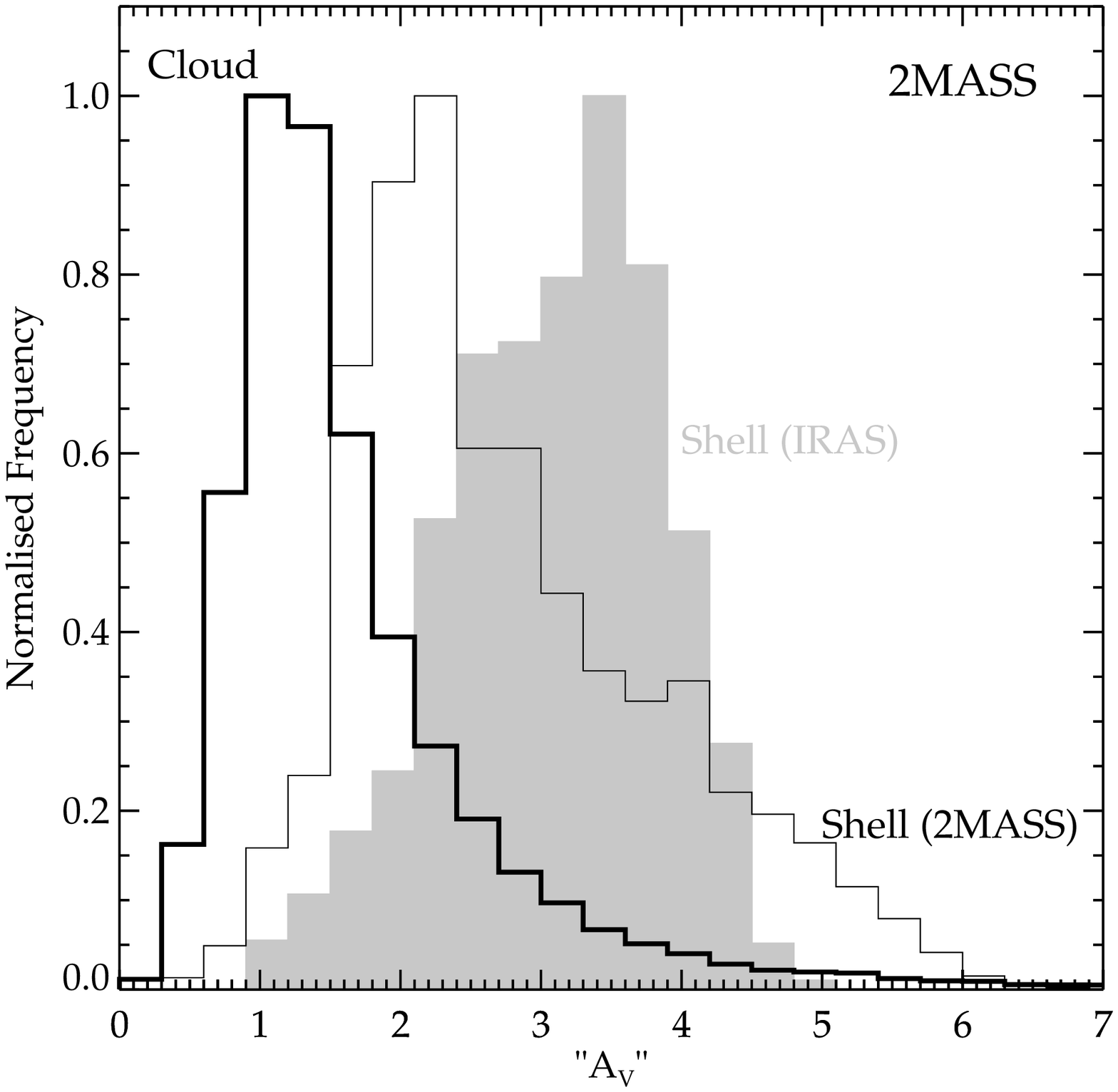}
\caption{Normalised 
histograms of extinction from IRAS and 2MASS/NICER.  In both panels,
the thick line represents the cloud component, and the grey filled
histogram shows the shell component as traced by IRAS (hatched region
in figure \ref{annulus}).  The thin line in the right histogram shows
the shell component as traced by 2MASS/NICER.  Note that small areas
containing the young clusters IC\,348 and NGC\,1333 were excluded in
both cases (see fig. \ref{annulus}), and that the histograms are
normalised due to the large difference in the number of points in the
two components (65000 points for the cloud versus 6800 for the shell
component).
\label{ext_hist}}
\end{figure}

Also shown in the right panel of figure \ref{ext_hist} is a histogram
of extinction as derived from 2MASS/NICER for the shell component.  If
we assume that the 2MASS/NICER derived extinction is a better estimate
of the ``true'' column-density of material towards any line-of-sight
then it appears that IRAS is systematically overestimating the
column-density of material in the shell.
  
Another way to look at this is figure
\ref{2mass_co_IRAS_scatter} where we show a scatter plot of column
density derived from IRAS against column density from 2MASS/NICER. A
solid line in the figure shows a linear least-squares fit to the data
and the dashed line indicates a 1:1 relation between the two
properties. Clearly, the shell component (red points) fall
preferentially above both the fitted line and the 1:1 line, again
indicating IRAS seems to be overestimating the column density in the
shell region.

\begin{figure}
\caption{IRAS column density vs. 2MASS extinction. The dashed line shows a 1:1 relation, and the solid line shows a fit to the data. The data clearly differ from the 1:1 line. Note that for clarity only 10\% of the green (cloud component) points are plotted, but the fit is heavily weighted by this component.
Symbols are the same as figure \ref{ext_temp_scatter}.
\label{2mass_co_IRAS_scatter}}
\end{figure}

The ``shell'' (red) and ``cloud'' (green) populations seen in figure
\ref{ext_temp_scatter} are less apparent when we make a similar plot
of temperature determined from IRAS against extinction determined from
2MASS/NICER, as is done in figure
\ref{2mass_iras_temp}. Although still warmer on average, the shell component
does not show a significant difference in column-density when compared
to the cloud component.

\begin{figure}
\caption{
Temperature from IRAS 60 and 100\,\um\ emission vs. extinction from
2MASS/NICER. Symbols are the same as figure \ref{ext_temp_scatter}.
\label{2mass_iras_temp}}
\end{figure}

We interpret the ``high contrast'' appearance of the shell on the IRAS
based column-density maps as caused by the presence of two different
dust populations along the line of sight, the cool component
associated with the molecular cloud complex, and a warmer component
due to the shell. This type of situation was discussed in an appendix
of
\citet{lwgb89}, who showed that the result of assuming a single 
dust temperature along a line of sight is an {\em underestimate} of
the total optical depth (and hence column density). However, in this
case we know the ``true'' column-density of shell material from the
2MASS/NICER map is in fact less than the column density we determine
from IRAS, i.e.  the line-of-sight dust distribution seems to cause an
{\em overestimate} of the 100\,\um\ opacity within the shell
component, as can be seen in the right panel of figure
\ref{ext_hist}. 

An alternative explanation is that our assumption of a uniform
dust-emissivity co-efficient for all of the region is incorrect. It is
possible that the grain-size distribution and/or composition of
the dust in the shell may be unusual, as it has been processed by the
radiation field of the massive star. 

The hypothesis of two cloud populations in the direction of Perseus is
not new. \citet{ut87} suggested the possibility of two clouds along
the line of sight to explain the presence of two distinct velocity
components seen in their CO map of Perseus, and in several papers
about photometric distances and extinctions toward field stars in the
directions of IC\,348, NGC\,1333 and the dark cloud
\objectname{Barnard 1}
\citep[][hereafter CS]{cernis90,cernis93,cernis03}, 
\citeauthor{cernis93}  and collaborators have found 
an apparent jump in extinction towards Perseus at a distance of
$\sim$140-160\,pc, placing some kind of absorbing material at this
distance. They attribute this material to an extension of material
from the Taurus molecular cloud complex along the line of sight to
Perseus (their suggested morphology is shown in figure 9 of
\citealt{cernis93}).

In a study of the optical polarization of background stars,
\citet[][hereafter GBMM]{gbmm90} found a bimodal distribution of 
polarisation angles in the Perseus region and attributed it to two
separate dust distributions along the line of sight, each associated
with a different magnetic field orientation. At the time, comparing
only to molecular-line maps, they could detect no meaningful spatial
distinction between the two polarisation
components. 

\citetalias{cernis03} have 18 stars in common with
\citetalias{gbmm90}, for 15 of which they were able to determine 
distances and extinctions, based on photometric spectral-typing.
These 15 stars' properties are summarised in table \ref{pol_tab} which
is reproduced in most part from \citetalias{cernis03}. We have split
the stars in the table into two groups, those with A$_V <$5 and those
with A$_V >$5.

From table \ref{pol_tab}, it is clear that the stars with {\em high}
polarisation percentage (bold entries in column 6 of the table) appear
almost exclusively in the {\em high} extinction ($\ga$0.5\,mag) group,
whilst stars with low polarisation percentage are more common at lower
extinction. The lower extinction stars also tend to be at nearer
distances.  \citetalias{cernis03} interpreted these results as further
evidence for their 140\,pc material, proposing that stars with low
polarisation are located between the 140\,pc material and the cloud
complex, while the more distant stars show high polarisation and high
extinction because they are behind two layers of absorbing material
(the 140\,pc material and the molecular cloud).

\begin{deluxetable}{ccccccc}
\tablecolumns{7}
\tablewidth{0pc}
\tablecaption{Distances, extinctions and polarisations of stars in Perseus, adapted from tables 1 and 3 of \citet{cernis03}.\label{pol_tab}}
\tablehead{
\colhead{(1)} & \colhead{(2)} &\colhead{(3)} &\colhead{(4)} &\colhead{(5)} &\colhead{(6)}&\colhead{(7)}\\
\colhead{ID\tablenotemark{a}}&\colhead{RA (J2000)} & \colhead{Dec (J2000)} &R &\colhead{A$_V$} & \colhead{P\tablenotemark{b}} & \colhead{$\Theta$}\\
\colhead{}&\colhead{hh:mm:ss} & \colhead{dd:mm.m} &pc &\colhead{mag} & \colhead{\%} & \colhead{deg}}
\startdata
\sidehead{Stars With A$_V <$0.7}
30&03:33:04 &30:34.5   &    268      &   0.21 &   0.51  &   87   \\
40&03:33:55 &31:10.0   &    310      &   0.29 &   {\bf1.24}  &   72   \\
51&03:33:41 &31:07.9   &    195      &   0.37  &  0.42  &   80   \\
71&03:35:41 &30:55.7   &    254      &   0.17  &  0.66  &   74   \\
72&03:35:52 &31:15.0   &    141      &   0.42  &  0.34  &   70   \\
88&03:36:36 &31:04.5   &    134      &   0.25  &  0.34  &   110  \\
\tableline
\sidehead{Stars With A$_V >$ 0.7}
15&03:31:54 & 30:55.8  & 700  &  {2.33}  &     {\bf1.14}  &   17   \\
24&03:32:44 & 30:30.3   & 300           &  {2.99}  &     0.21  &   114  \\
31&03:33:10 & 30:50.3   & 860           &  {2.18}  &     0.43  &   143  \\
74&03:35:55 & 31:33.8   & 440           &  {1.38}  &     {\bf1.46}  &   146  \\
92&03:36:42 & 31:31.1    & 1430          &  0.71  &     {\bf1.12}  &   156  \\
111&03:37:33& 30:58.9  & 264           &  {2.47}  &     {\bf4.69}  &   145  \\
112&03:37:45& 31:07.0 & 5200  &  {1.24}  &     {\bf5.06}  &   152  \\
124&03:38:24& 31:12.0  & 245           &  {3.56}  &     {\bf9.14}  &   150  \\
125&03:38:36& 31:56.5  & 340           &  {1.66}  &     {\bf1.23}  &   137  \\
HD\,278942&03:39:55  &31:55:33.2 & 207\tablenotemark{c} &{\bf7.4}\tablenotemark{d,\rm{e}} & {\bf5.9}\tablenotemark{d} & 154\tablenotemark{d}\\
\enddata
\tablenotetext{a}{Notation of \citetalias{cernis03}}
\tablenotetext{b}{Entries shown in {\bf bold face} have polarisation percentage $>$ 1\%.}
\tablenotetext{c}{Hipparcos \citep{hipparcos}.}
\tablenotetext{d}{\citet{ander2000}.}
\tablenotetext{e}{Contains a significant circumstellar 
component \citep{ander2000}.}
\end{deluxetable}

In figure \ref{pol} we show polarization vectors from
\citetalias{gbmm90} overlaid on our IRAS column-density and 
\xco\ maps. The weaker
polarisation component identified by
\citetalias[red vectors;][]{gbmm90} is clearly 
aligned along the major axis of the molecular cloud as stated in that
paper, but on comparison with our new column-density maps it now
becomes evident that much of the high polarisation-percentage
component is aligned with the warm shell and associated
arm to the west.

\begin{figure}
\caption{IRAS colum density (left) and extinction from 2MASS (right)
overlaid with polarisation vectors from \citetalias{gbmm90}.  Blue
vectors have polarisation strength, P $>$1.2\% and red vectors have P
$<$1.2\%). The stronger polarisation component (blue) appears
aligned with the warm dust emission seen in the IRAS map,
while the weaker component mostly lies along the chain of molecular
clouds, most evident in the 2MASS extinction map.
\label{pol}}
\end{figure}

In fact, the polarisation strength ($\la$1.2\%) measured for the
weaker component is typical for nearby molecular clouds
\citep[e.g.][]{gbmm90,bhatt04}. We therefore propose that the weaker component 
is due to magnetic fields within the molecular cloud material, whilst
the stronger component, is due to a swept-up field associated with the
shell material. This is consistent with \citetalias{cernis03}'s
measurements, as the distances to the low-polarisation stars (non-bold
entries in column 6 of table \ref{pol_tab}) place them close to the
distance of the molecular cloud complex, whilst the
higher-polarisation stars (bold entries in column 6 of the table) are
generally more distant than 300\,pc. \citet{ander2000} measured a
polarisation of 5.9$\pm$0.03\% at a position angle of
154\fdg0$\pm$0\fdg2 towards HD\,278942, also consistent with its
polarisation being associated with the shell in our scenario (although
we note that some fraction of the polarisation is thought to be
circumstellar).

Although we do not know its distance, the shell cannot be the source of
\citetalias{cernis03}'s proposed 140\,pc material, as it must lie {\em behind} 
the molecular cloud complex.  We know this, because in figure 
\ref{halpha_2mass}, which shows contours of extinction from 2MASS/NICER
overlaid on the H$\alpha$ image, a finger of extinction (contours), is
exactly co-incident with a dark ``shadow'' to the southeast in the
H$\alpha$ emission. The same shadow can be seen in the 1907 photograph
in Barnard's Atlas \citep{barnard}, against a circular shaped
nebulosity which is located within the circumference of the
shell. This would place the shell behind at least the eastern end of
the Perseus molecular cloud complex, and therefore more distant than
260$\pm$20\,pc, the distance to IC\,348 \citep{cernis93} which is
embedded in that portion of the molecular cloud\footnote{This makes
either the distance determination to HD\,278942, or its association
with the shell somewhat problematic, as its Hipparcos distance of
207$\pm$52\,pc
\citep{hipparcos,ander2000} would place it on the near side of the cloud 
containing IC\,348, while its high reddening and polarisation would
suggest it is behind a significant column of dense material.}. We also
do not believe that the shell could be far behind the molecular cloud
complex, as several of the more highly polarised stars in table
\ref{pol_tab} have distances $\la$300\,pc.

\begin{figure}
\caption{
H$\alpha$ emission \citep{fink04,vtss}, overlaid with contours of
extinction from 2MASS/NICER. The extinction from the molecular clouds
obscures the southwest quadrant of the ionized emission,
indicating that the H$\alpha$ emission, and hence the shell, is behind
the cloud complex.
\label{halpha_2mass}}
\end{figure}

\subsection{Is the shell impacting on the molecular cloud?}
\label{impact}

The final piece of the puzzle of the Perseus shell is whether or not
it is interacting with the molecular cloud. In earlier sections we
have argued that it is likely to be located behind, but still close to
the cloud complex. The most compelling evidence for an interaction is
given by a comparison of the MSX 8\,\um\ image with extinction from
2MASS as was shown in figure \ref{msx}.  The shell is clearly
detected in 8\,\um\ emission, and in particular has three bright
``knots'' of emission\footnote{see also \citealt{kraemer} for images
at all 4 MSX bands.}, labelled A, B and C in figure \ref{msx}). An
extinction of A$_V \ga$ 100 is required to make the cloud opaque at
8\,\um, so the bright knots are likely to be real and not just a
result of varying extinction in front of the emission.  In fact, the
three bright knots are spatially coincident with slight enhancements
in the extinction, as indicated by the contours on figure \ref{msx}.
Emission in the 8\,\um\ band of MSX has been modelled as a combination
of grey-body dust emission and emission from the unidentified infrared
bands (UIBs), usually attributed to polycyclic aromatic hydrocarbons
(PAHs) excited in shocks \citep[e.g.][]{go}, and hence the spatial
coincidence between the 8\um\ knots and the extinction enhancements
are most easily explained by a layer of swept up material in front of
a shock, providing compelling evidence for an interaction. Further
evidence is seen in the \xco\ emission, where the spectrum at ``knot
C'' shows a strong second component at 5\,\kms\ (figure \ref{co_spec})
well separated spatially from emission due to the molecular clouds at
those velocities (figure \ref{co_chann}).  The second component is
blueshifted with respect to the ``ambient'' molecular gas, consistent
with an interaction with the front side of an expanding shell.

\begin{figure}
\caption{A ``sampler'' of \xco\ spectra from the Perseus region.
The bottom spectrum shows the mean spectrum of all positions where
\xco\ was detected at the 5$\sigma$ level within the boundaries shown
in figure \ref{co}. This spectrum is multiplied by a factor of
4. Shown above the average spectrum are three spectra from the
positions of the 8\um\ knots labelled A, B and C in figure
\ref{msx}. The component at 5\,\kms\ in knot C is due to an
unusual-looking small feature which appears unconnected either
spatially or in velocity to the rest of the cloud complex (also see
figure \ref{co_chann}), and hence may be due to an interation between
the shell and the cloud complex. At the top, three spectra from the
position of peak \xco\ emission in the known star-forming clumps 
NGC\,1333 ($\alpha_{J2000}$=03$^{\rm h}$29$^{\rm m}$09\fs5, 
$\delta_{J2000}$=+31\degrees21$^{\rm m}$45$^{\rm s}$), 
IC\,348 ($\alpha_{J2000}$=03$^{\rm h}$44$^{\rm m}$03\fs9, 
$\delta_{J2000}$=+32\degrees04$^{\rm m}$23$^{\rm s}$)
and B\,1 ($\alpha_{J2000}$=03$^{\rm h}$34$^{\rm m}$36\fs0, 
$\delta_{J2000}$=+31\degrees23$^{\rm m}$34$^{\rm s}$)
are shown for comparison. The dashed lines show the central
velocity of the two components of knot C.
\label{co_spec}}
\end{figure}
\begin{figure}
\caption{
The bottom panel shows total integrated \xco\ intensity near the
shell. The cross indicates the position of HD\,278942. The three
triangles show the positions of the three 8\um\ knots which were also
indicated in figure \ref{msx} (A, B and C, from left to right).  The
top panel shows channel maps in intervals of 2\,\kms\ of the same
region. Velocities are given in the top right corner of each
panel. The bright feature at the center of the 4--6\,\kms\ channel
(top right) is due to the 5\,\kms\ component seen in the spectrum at
the position of 8\,\um\ knot 'C' (figure
\ref{co_spec}), but is not physically connected to the emission
associated with the molecular clouds at similar velocities.
\label{co_chann}}
\end{figure}

CO spectra in the Perseus region are extremely complex and confused,
with a velocity gradient of almost 10\,\kms across the cloud and
multiply peaked, optically thick lines throughout. We have attempted
to make \co\ observations of the top half the shell, where
contamination from the molecular clouds should be low in order to
perform a more detailed kinematical analysis of the shell, but as can
be seen from the 2MASS/NICER extinction data, the column density there
is low, and we have been unable to detect \co\ at sufficient
sensitivity to enable such an analysis.

By assuming a gas-to-dust ratio $\frac{N_H}{A_V}$, we
can calculate the mass of material in the shell from the 2MASS
extinction map, using the relation presented by \citet{dickman}:
\begin{equation}
M = (\alpha d)^2 \mu \frac{N_H}{A_V}\sum_{i} A_V(i),
\end{equation}
where $\alpha$ is the angular size of a map pixel, $d$ is the distance
(300\,pc), $\mu$ is the mean molecular weight corrected for helium
abundance and $\frac{N_H}{A_V}=1.87\times 10^{21}$cm$^{-2}$mag$^{-1}$
\citep{sm79}, although this number could be higher by a factor of
$\sim$2, \citet[see e.g.][]{gh94}.
Due to the contamination from cloud emission in the lower half of the
ring, we calculated the mass in the top half of the shell and scaled
this by a factor of two to estimate a total mass of the material in
the shell of 760\Msolar. If we assume we are seeing the limb
brightened edge of a complete spherical shell then the total mass,
$M_{TOT}$ such a shell would contain is given by:
\begin{equation}
M_{TOT} = M_{RING} \left\{\frac{R_2^3 - R_1^3}{\left[R_2^2-R_1^2\right]^{3/2}}\right\},
\end{equation}
where $M_{RING}$ is the mass enclosed within the 2-dimensional annulus
indicated in figure \ref{annulus} and $R_1$ and $R_2$ are the inner
and outer radius of that annulus respectively. In this case $R_1 =
2.05$ and $R_2 = 4.29$\,pc (d=300\,pc), leading to a total mass of the
shell of 1000\,\Msolar. Assuming a typical B-star lifetime of 8\,Myr
and a radius of 4.3\,pc gives a time-averaged expansion velocity of
0.5\,\kms\ for the shell, and a total outward momentum of 520\,\Msolar
\kms, much less than 10$^5$\,\Msolar
\kms\ we would estimate based on the properties of the central star
(see equation 18 of \citealt{matzner}). This would suggest that the
shell is in the process of merging with the background gas, in which
case it is surprising to see such a coherent ring. This
discrepancy could be explained if the shell was initally expanding
more quickly into a higher density material. For instance, the same
diameter could be reached in 1\,Myr with an expansion velocity of
4\,\kms.

\section{Summary}

By combining our reanalysis of ISSA data \citep{schnee_iras} with data
from the COMPLETE Survey of Star-Forming Regions \citep{data} and
archival H$\alpha$ and mid-IR observations, we have made a detailed
study of the warm dust shell in Perseus which was first reported by
\citet{ps89}.

\begin{enumerate}

\item	
A comparison of the temperature and extinction derived from IRAS 60 and
100\um\ emission shows two distinct dust populations in the Perseus
region. The two populations are clearly spatially separated, with the
warmer population being associated with the shell.

\item
The shell is also detected, but at much lower contrast, in an
extinction map of the same region constructed by applying the NICER
algorithm \citep{la01} to 2MASS \citep{data}. This indicates that the
shell is not as significant a column-density feature as one would
infer from the IRAS map. It dominates the IRAS column-density map
because it is warm.

\item
The shell is spatially coincident with, and hence may be the source of
the high-polarisation component of the bimodal distribution of
polarisations found in the region by
\citet{gbmm90}.  If this is the case, then this would be the first
observation of a swept-up magnetic field on $\sim$10pc scales.

\item
The combination of IRAS, MSX, 2MASS/NICER and \xco\ data suggests that
the shell is located behind the molecular cloud containing the
star-forming cluster IC\,348, and there is significant circumstantial
evidence that it is interacting with the cloud complex.

\item 
We reiterate previous conclusions \citep{ag99a,ag99b,cbls,schnee_iras}
that care should be taken when using IRAS to determine the extinction
in the direction of complex regions such as star-forming molecular
clouds, as variations in dust temperature along the line of sight can
introduce substantial bias toward warm dust features.
\end{enumerate}

\acknowledgments
This research has made use of the NASA/IPAC Infrared Science Archive,
which is operated by the Jet Propulsion Laboratory, California
Institute of Technology, under contract with the National Aeronautics
and Space Administration; and data products from the Two Micron All
Sky Survey, which is a joint project of the University of
Massachusetts and the Infrared Processing and Analysis
Center/California Institute of Technology, funded by the National
Aeronautics and Space Administration and the National Science
Foundation. We acknowledge the use of NASA's {\em SkyView} facility
(http://skyview.gsfc.nasa.gov) located at NASA's Goddard Space Flight
Center. FCRAO is supported by NSF Grant AST 02-28993.

\bibliographystyle{apj}
\bibliography{refs}


\end{document}